\documentstyle[12pt,psfig,rotate]{article}
\newcommand{\draft}{
        \renewcommand{\baselinestretch}{1.0}%
        \small\normalsize%
}
\newcommand{\final}{
        \renewcommand{\baselinestretch}{2.0}%
        \small\normalsize%
}
\draft
\begin{document}
\title{\bf On Period and Burst Histories of AXPs and SGRs and the 
Possible Evolution of these Objects on the P-\.{P} Diagram} 
\author{Oktay H. 
Guseinov$\sp{1,2}$ \thanks{e-mail:huseyin@pascal.sci.akdeniz.edu.tr}
, Efe Yazgan$\sp2$ 
\thanks{email:yazgan@astroa.physics.metu.edu.tr} \\ \\
A\c{s}k\i n Ankay$\sp2$ 
\thanks{e-mail:askin@astroa.physics.metu.edu.tr}
, Sevin\c{c} Tagieva$\sp3$
\thanks{email:msalima@lan.ab.az} \\ \\
{$\sp1$Akdeniz University, Department of Physics,} \\ 
{Antalya, Turkey} \\
{$\sp2$Middle East Technical University, Department of Physics,} \\
{Ankara 06531, Turkey} \\ 
{$\sp3$Academy of Science, Physics Institute,} \\
{Baku 370143, Azerbaijan Republic} \\ \\ \\ \\ \\ \\ 
}

\maketitle
\final
\newpage
\begin{abstract}
\noindent
In this paper, timing data for all of the anomalous X-ray pulsars and 
soft gamma repeaters are compiled. Timing properties of these objects are 
investigated. The effect of bursts of soft gamma repeaters on their 
period history is investigated.
The P-\.{P} diagram for pulsars, X-ray binaries, anomalous X-ray pulsars, 
soft gamma repeaters and dim radio quiet netron stars is constructed.
The possible evolutionary tracks for anomalous X-ray pulsars, soft gamma 
repeaters and dim radio quiet netron stars are examined. 
\\ \\ KEY WORDS: AXP, SGR, DRQNS, MAGNETAR \end{abstract}

\clearpage
\parindent=0.2in
\section{Introduction} 
The new kind of neutron stars (NS); so-called anomalous X-ray pulsars (AXPs)
and soft gamma repeaters (SGRs) are characterized by their rapid spin 
down $[1]$.
To explain this rapid spin-down, and the other characteristics of 
these objects such as the low rotational energy loss, two fundamentally 
different
models are proposed: accretion from a fall-back disk model and the magnetar
model.
Accretion from a fall-back disk model describes these objects as
systems with disks which have just finished the so-called propeller phase
$[2]$.
However, it is shown that accretion from a
fall-back disk model has too many difficulties in the explanation of AXPs 
$[3]$. An expansion of this model might include the effect of propeller
and accretion at the same time. Such a model will be considerably 
successful in the explanation of properties of AXPs, but not SGRs $[3]$.

To explain the strong gamma-ray bursts and the persistent X-ray 
radiation, 
it is proposed that the magnetic field of SGRs and AXPs 
decay in $\sim$10$^4$-10$^5$ yr. Moreover,
the P-\.{P} values and the strength of the giant gamma ray bursts 
suggest very high magnetic fields for these 
objects. According to the magnetar model, AXPs and SGRs
have magnetic fields of $10^{14}-10^{15}$ G $[4]$.  
Magnetar models are more attractive because they utilize new physical 
phenomena trying to explain the AXP and SGR properties together 
simultaneously. Since accretion theory is very well developed and well 
known, it is much easier to criticize the accretion models proposed for AXPs 
and SGRs.  
This is the first of a series of papers showing the problematic and 
good aspects of the magnetar model depending on the observational data.

Three AXPs are connected to supernova remnants (SNRs) $[5,6]$ and one of the 
SGRs, namely 
SGR 0526-66, is possibly genetically connected to SNR N49 in the Large 
Magellanic Cloud [6]. 
About 21
radio pulsars (PSRs) are connected to galactic SNRs $[7,8,9,10]$
with high
probability. Two of these AXPs (AXP 
1E1841-045 and AXP 1E2259+586 which are
located in SNRs G27.4+0.0 and G109.1-1.0 respectively) are not farther   
than 7 kpc from the Sun. Observational data show that in that volume the 
number of SNRs is not less than 100.  
Contrary to PSRs, AXPs can be detected at such distances relatively easily.
From the above arguments, it follows that the birth rate of AXPs is at least 
50 times smaller than the birth rate of SNRs.

Some recently
found PSRs have magnetic fields of B=$10^{13}-10^{14}$ G  $[11]$.
We do not yet know the properties of these PSRs in detail,
but young NSs with B$<10^{14}$ G (having ages less than 10$^5$ years) in 
general,  
have very low X-ray luminosities and do not have $\gamma$ ray burst
activity. However, observed NS-magnetars 
always have L$_x > 10^{34}$ erg/s and
they sometimes have strong $\gamma$ ray burst properties.
Whether this statement about NS-magnetars is true in general 
is not known since number of data are few.

AXPs and SGRs are distant objects.
Therefore it is very difficult to obtain trustworthy and
sensitive enough observational data from these objects. This is also 
true for the SNRs which might be genetically connected with them. 
Moreover, the number of AXPs and SGRs are so small 
to perform statistical analysis. This situation cannot considerably
change in the future. The direct estimate of the magnetic field 
strength of these objects from cyclotron line detection is more 
important. It is such a pity that we do not yet detected such a cyclotron 
line. Therefore for the investigation of origin and
physical nature of these very important objects, it is necessary to study
also neutron stars which demonstrate other types of properties. 
Young radio pulsars (PSRs) and single radio quiet neutron stars (DRQNS)
which are observed as dim point X-ray sources will provide valuable 
information. In this paper we 
concentrate our attention on the timing properties of AXPs and SGRs.
Also, we want to understand the  
possible evolution of AXPs and SGRs on the P-\.{P} diagram, qualitatively. 

\section{The Period Histories of AXPs and SGRs}
\subsection{The Period Histories of AXPs}
Figures 1-7 which are constructed 
using the data in Tables 1-7 display the period histories of AXPs 
and SGRs. We have investigated, how braking index of 
AXPs and SGRs change using these data. 

\.{P} of AXP
1E2259+586 and AXP 1841-045 had been constant for a long time from 1978
to 2000 and 1987 to 1999 respectively, as though braking index on the 
average were constant.
Average value of \.{P} for AXP 1E2259+586 is $\sim$5.4$\times
10^{-13}$ s/s. For AXP 1E1841-045 average value of \.{P} is
$\sim$4.1$\times 10^{-11}$ s/s (see Figures 1 and 2 and Tables 1 and 2). 
These objects are connected with SNRs, so
they are very young. Deviation of observational values of periods  
from constant \.{P} line is related to noise, as well as observational errors.  
The absolute values of deviation of the 
period from constant 
\.{P} line for AXP 1E2259+586 is not more than the deviation for 
AXP 1E1841-045 (we want to note that the deviation for AXP 1E2259+586 
seems to be more than the deviation for AXP 1E1841-045 since the $scale$ 
of the figures are not the same).

Braking of AXP 1048.1-5937, which is not connected with an SNR, had 
sharply increased (braking index has decreased) between the years 
1990-1998. 
The braking of AXP 1E1048.1-5937 
from 1979 to 
2000 changed considerably ($\sim$ 2 times). From 1979 to 1988, from 1988 to
1994 and from 1996 to 2000 average values of \.{P} were about 1.6$\times
10^{-11}$, 2.6$\times 10^{-11}$, and 1.9$\times 10^{-11}$ s/s
respectively. Average value of \.{P} was roughly 3$\times 10^{-11}$
s/s for the 12 years period from 1988 to 2000. 
From the fact that the average is higher than the values in the three 
intervals quoted above 
is a result of the period jump between 1994-1996 
(see Figure 3 and Table 3). 

What is the reason for such huge changes of braking index? 
As seen from Figures 1-7 and Tables 1-7 there are both sharp increase and 
decrease of periods of AXPs and SGRs. 
To change the period, moment of inertia must be changed due to burst or 
glitch and/or changing magnetic interaction with the surrounding matter 
during the bursts due to plasma ejection.
If these drastic changes have connection
with bursts in deep levels of the atmosphere, then AXP 1E1048.1-5937 must 
have a high  
temperature. As seen from Table 8,  it has large but not the highest 
blackbody temperature in 0.1-2.4 keV band 
among all AXPs. Its luminosity also is not the highest one in 2-10 keV 
band (Table 8). This AXP has the largest value of \.{E}, but also the 
largest 
value of deviation (when scales of figures are taken into account) from 
the constant \.{P} line among all the other 
AXPs, therefore stronger energetic processes must take place.
In the frame of the magnetar model, 
it is difficult to understand the fact 
that AXP1E1048.1-5937 does not have the largest temperature 
and luminosity.

AXP 1E1048.1-5937 like AXP 1E2259+586 has been observed for 20
years, and AXP 1E1841-045 has been observed since 1987. However, no gamma ray
burst have been detected from these objects. But these are distant 
objects and the observations may not be long enough. Therefore, we may 
say that AXP 1E1841-045 (even though it has   
constant \.{P}) and AXP 1E1048.1-5937 will show burst properties in the
future. 

Braking of AXP 4U0142-614 had increased in the years 1995-1998
(see Figure 4 and Table 4). For this AXP average value of \.{P} between 
1979-1993
is $\sim$1.9 $10^{-12}$ s/s and after 1993, $\sim$3 $10^{-12}$ 
s/s. Average value of \.{P} between 1984 and 1998 is $\sim$2.4 $10^{-12}$ 
s/s. 

A strong glitch was observed from AXP 1RX J170849-4009 in 1999 
$[12]$ similar to the glitches 
for the
Vela PSR $[13]$. This probably have influence on the 
the period history and \.{P}. But for this AXP, there are not enough 
period measurements (see Figure 5 and Table 5). This AXP, which do not 
have a connection with an SNR, is discovered in September, 1996 $[14]$.   
Period of AXP J170849-4009 is P=11.0 s and \.{P}=2.25$\times 
10^{-11}$ s/s $[15]$. From Figure 5 we found 
\.{P}=2.98$\times 10^{-11}$ s/s. 
At the moment, it is 
difficult to talk about the character of period history for this source.

The X-ray pulsations of AXP J1845.0-0300 has been found in 1998 $[16]$
and after that no more period value is found for this source, so
period change of this source cannot be determined directly now. But with 
an indirect approach we can determine the lower limit for \.{P} as follows. 
This AXP is connected with SNR G29.6+0.1 which has age $<$8$\times
10^{3}$ years $[17,18,19]$. 
Equating the age of the SNR to characteristic age $\tau$ and taking n=3 
we estimate 
\.{P}$>$1.3$\times 10^{-12}$ s/s. 
But if the
magnetic field of this source is decaying then
n$>$3. This lowers the value of \.{P} and increases $\tau$. Probably 
\.{P} value is about (1.3-2)$\times 10^{-12}$ s/s. 
We have used this value in the P-\.{P} diagram (Figure 8).

\subsection{The Period Histories of SGRs}
After three
strong bursts in 1979 March 5 with intensities 
$\sim$(1-3.5)$\times 10^{-6}$ erg/cm$^2$ s $[20]$, SGR 1900+14 
became
active again after quiescence in 1992 with new strong 
bursts. But intensities of these bursts were several orders of magnitude 
smaller; 
about (3-6.6)$\times 10^{-8}$ erg/cm$^2$ s $[21]$.
Braking of SGR 1900+14 had increased in the period 1998 May-1999 January
twice, after 63 bursts $[22]$. Intensities of these bursts (see Table 8) 
were even larger than the ones in 1979 and raised up to $>$2.1$\times 
10^{-4}$ erg/cm$^2$s $[23,24]$. For the 
strongest burst which occurred in 1998 August 27 the
burst energy was $\sim$5$\times 10^{42}$ erg $[23]$. 
In result of these bursts between 1998 May - 1999 March (Table 6 
and Figure 6) 
value of \.{P} considerably increased.  It is very interesting that 
there were large increase 
in period also between 1999 March and 2001 February, 
because next burst occurred in 2001 April 18 
$[25]$. The reasons might be the plasma 
ejection
and interaction with the surrounding matter. This burst had an 
intensity of 1.1$\times 10^{-5}$ erg/cm$^2$-s and
a flux of 1.4$\times 10^{-6}$ erg/cm$^2$-s. The total burst energy 
was $\sim$7$\times 10^{40}$ erg in 2001 July 2 $[26]$. 
Average value of \.{P} in the years from 1995 to 1999 is $\sim$6 
$10^{-11}$ s/s and
$\sim$1.6 $10^{-10}$ s/s for the years between 1998-2001. The average 
value of \.{P} for the years between 1995-2001 is $\sim$1 $10^{-10}$ s/s. 
Without any doubt for this SGR, \.{P} should fall to several times 
lower values in the future.

SGR 1806-20 had been active between 1978-1987 and especially in
1983 $[27]$ when the peak flux was $>$10$^{-5}$ erg/cm$^2$s 
$[28]$. As 
seen from Tables 9 and 7 and Figure 7, this SGR has several active periods 
between the years 1993-2001.
\.{P} of
SGR1806-20 had been constant practically, between 1993-1999, and equal to 
$\sim$6 $10^{-11}$ s/s. According to [29] the period 
had increased a little during the burst in
1999 January, 22 and after that it had decreased. 
According to [30]
the energy for 1993-1999 bursts were 
$\sim$10$^{39}-10^{41}$ erg. 
In this period of time,  we see the influence of these bursts on the period 
history, 
similar to the effect of bursts on the period history of SGR 1900+14. From 
Figure 7, we see that \.{P} increases 
sharply after the 1999 burst activity. Above, we have seen a similar 
behavior for SGR 1900+14.  

The instantaneous   
spin-down rate for SGR 1806-20 at a single epoch was higher by a factor of 
1.5 than 
the   global trend [29]. Between the RXTE measurement in May 1999 [29] 
 and the CHANDRA measurement in August 2000 [31] 
, the average spin-down rate 
is a factor of 2 higher than the largest \.{P} previously measured for
this source. 
Several days before (2000 August 15) the CHANDRA
observation, there were moderate activity for SGR 1806-20; but this was too 
close to 
the CHANDRA observation to have affected the spin period significantly. 
Without any doubt the sharp increase of \.{P} of SGRs are related with 
burst activity. But as seen from period histories of SGRs, the the 
time-scale of increase 
of P and \.{P} values (decrease of braking index) are not more than 
1-3 years. Therefore, the process decreasing the braking index 
due to 
the ejected plasma finishes before the plasma reaches a distance  
$<$ 1 pc from the NS. 

\subsection{The Braking Noise in the Periods of AXPs, SGRs and PSRs}
Magnetic decay can create noise in the period history of NSs. Magnetic decay 
in the long run decreases \.{P} and makes n$>$3.
However, braking index is the result of all the processes that brake the
star's rotation. Of course, magnetic decay decreases the amount of magnetic 
dipole
radiation but in this process other effects can increase braking. For  
example, ultra-relativistic particle winds decrease the value of n.
Increase in \.{P} was the highest for SGR 1900+14 (during and after the 
1998 bursts) and SGR 1806-20 (after the 1999 bursts).
But what causes the strong and irregular change in periods for AXP 
1E1048.1-5937, 4U0142+61 and possibly RXSJ170849-4009 which are not located 
in SNRs and do not have pulsar wind nebulae (PWN)?

At birth, what brakes the rotation of the PSR might not only be the 
magnetic dipole radiation but also something
other than the magnetic-dipole radiation. Then,  the
braking index of the PSR can be different from 3. The braking index of PSRs 
being close to 3 shows that the dominant mechanism is the magnetic dipole
radiation [13]. On the other hand, the characteristic 
times of different braking mechanisms are different so that we expect such 
situations: 

1. The characteristic times of period change in result of action of other 
mechanisms are shorter compared to that of magnetic dipole radiation. 
The effects of such other mechanisms are not observable for newly born 
PSRs that it is impossible to observe these effects for older PSRs.

2. Assume that the characteristic time for the other mechanism is higher
than that of magnetic dipole radiation for young PSRs with strong 
magnetic fields. In this case the other mechanism
can show itself in old PSRs and in PSRs with small magnetic fields 
(effect of dipole is weak so that the effect of the other mechanism can 
be seen).  

Evidently, any mechanism other than magnetic
dipole radiation has much more noise than magnetic dipole radiation and 
braking index is different from 3. On
the other hand, since the rate of rotational energy loss decreases very
fast, noise should increase as the period increases (because 
the other mechanism produces noise). However, we see just the 
opposite for PSRs.

So if there exist some other mechanisms, they must show themselves only
in young PSRs and the characteristic times must be shorter than that of  
magnetic dipole radiation mechanism, because, only young PSRs with large 
value of magnetic fields have braking index smaller than 3 and large 
values of noise for comparison with other PSRs. This must be true for 
all types of single neutron stars (SGRs, AXPs and DRQNSs). For AXPs and 
SGRs 
another braking mechanism exist and value of
\.{E} is low, so there is large period noise. In the frame of the 
magnetar model, existence of noise for AXPs shows that there are 
continuous magnetic decay and burst activity for AXPs.

The PSR with the least braking index is the Vela PSR (PSR J0835-4510),  
n=1.4$\pm$0.2 [33]. This is natural since    
this PSR has the strongest glitches. 
Of course smallness of n for Vela PSR is also related with its PWN.
The braking index of PSRs J0534+2200, J0540-6919 (which are not detected 
at 1400 MHz as belonging to Magellanic Cloud), J1513-5908 and J1119-6127 are
respectively, 2.51$\pm$0.01 [34], 2.2$\pm$0.1 [35]  
, 2.837$\pm$0.001 [36] and 2.9$\pm$0.05 [11]
 and all of them have characteristic ages $\tau < 10^4$ years
since these PSRs have genetic connections with SNRs.
Data about these PSRs and SNRs connected with them are given in Table 10. 

As seen from Table 10, braking index of PSRs do not depend on the value 
of the  
pulsed radiation (radiation related to curvature radiation) and 
properties of 
NSs. They depend on the amount of magnetic dipole radiation (related to 
\.{E}) (the reasons that n is small are strong PWN and glitches). 

The 
evolution for radio PSRs are on the average with n=3 and noise for radio 
PSRs are  very 
low; so dominant mechanism is the magnetic dipole radiation.
Therefore magneto-dipole radiation in general determines the value of 
\.{P} and \.{E}. Evidently, existence of 
PWN require additional rotational energy loss. 
The 
deviation of the value of the braking index from n=3 can not be 
determined by using only the size and the amount of radiation of PWN. But we 
see the dependence of these deviations on the morphological type of SNRs and 
on the properties 
of PWN. In all cases when PWN exist, value of n$<$3. PSRs with strong 
PWN in Crab and N157B have small deviation of n from 3 than Vela because 
of the strong glitches of the Vela PSR. Naturally single NSs 
without glitches, PWN, accretion or propeller must have n=3 
in result of pure magnetic dipole radiation. AXPs and SGRs do not 
have strong PWN nor do they demonstrate glitches similar to Vela pulsar 
(with the exception AXP RXSJ170849-4009, or we cannot observe glitches as 
periods have large noises). Moreover, if 
they are not accreting stars, then they must have in average braking index 
more than 3 because of magnetic field decay. But for SGRs and AXP 
1E1048.1-5937 we see considerable change of \.{P} and braking index. It 
shows that often the braking of NSs which are related to other mechanisms 
may be larger than the effect of magneto-dipole radiation braking. 
Therefore, often, the braking index may be $<$3 also.

\subsection{Possible Evolutionary Track of AXPs and SGRs on the 
P-\.{P} Diagram}
AXP 1E1048.1-5937 is similar to SGR 1900+14 and SGR 1806-20 according to 
their small L$_x$/\.{E} values (see Figure 9 and Table 8) and according 
to their location on the P-\.{P} diagram (see Figure 8). 
These three objects have also large noise in period. The small value of the 
ratio L$_x$/\.{E} is the result of the large value of \.{E}. In Figure 9, 
also the L$_x$/\.{E} values of DRQNSs given in Table 11,  
young PSRs ($<$10$^5$ years) radiating also in X-ray band, are shown 
[37] (values of L$_x$/\.{E} for these PSRs are 
given in 
Table 12). Also in this Figure there are two old PSRs, namely, PSR 
J0826+2637 and PSR J0953+0755 having ages of 5 10$^6$ and 2 10$^7$ years. 
There are also other PSRs with similar ages having total X-ray radiation 
about 
10$^{30}$ erg/s. But DRQNS RXJ0720.4-3125 have an X-ray luminosity of 1.7 
10$^{31}$ erg/s in 0.1-2.4 keV band at distance 0.3 kpc [38,39]
. This source have $\tau$=(2.5-3.6) 10$^6$ years 
(see Table 11). This value is several times larger than that of ordinary 
PSRs  with similar ages. It may be the result of decreased 
cooling time with increased magnetic field and smaller mass (since 
this source has B$\sim$3 10$^{13}$ G, see Figure 8)

In Figure 8, other than AXPs and SGRs, also P-\.{P} values for young PSRs 
with characteristic 
ages $\tau<$10$^5$ years, DRQNSs (from Table 11) and low mass X-ray 
binaries (from Table 13) are shown.
As seen from the P-\.{P} diagram, AXPs have, on the average, up to 2 
orders of 
magnitude  lower \.{E} values compared with SGRs and also AXPs are about 4 
order of magnitude younger than PSRs on the average.

The radio pulsars PSR J1119-6127 ($\tau$=1.6 kyr), PSR
J1814-1744 ($\tau$=85 kyr), PSR J1726-3530 ($\tau$=14.5 kyr) and 
PSRJ1734-3333 ($\tau$=8.1 kyr) have  magnetic field strength of
4.1$\times 10^{13}$, 5.5$\times 10^{13}$, 3.98$\times 10^{13}$ and 
5.4$\times 10^{13}$ Gauss, respectively [11,40].
These values are close to the value for AXP 1E2259+586 (B=5.9$\times 
10^{13}$ Gauss). 
Moreover, for the very young PSR J1119-6127 ($\sim$1.6 kyr) the upper 
limit for the X-ray luminosity in 0.1-2.4 keV and 2-10
keV bands is about an order of magnitude less than the X-ray luminosity (in
the same bands) of
AXP 1E2259+586 [41,11]. All of these 
PSRs are distant
objects. PSR J1119-6127 (l=292.2), J1814-1744 (l=13), J1726-3530
(l=352.4) and J1734-333 (l=354.8) respectively have distance values of 7.5, 
10.2, 10 and 7.4 kpc. PSR
J1119-6127 is connected with SNR G292.2-0.5 with very high probability
[42,7,43]. 
PSR J1726-3530 is probably connected with SNR G 352.2-0.1 [43]
. Both of these SNRs are shell type, similar to SNRs which are 
connected with AXPs. These PSRs and also AXPs do not have strong PWN which 
may be easily observed at 5-10 kpc distances. 
However, these PSRs do not have large noise and they 
have low X-ray luminosity. This shows that, as we increase the magnetic 
field of 
normal PSRs, X-ray luminosity do not increase nor do the noise. The 
non-increase of noise seems normal since \.{E} is high, but at least they 
should have a little higher noise than other PSRs as they have closer 
magnetic 
fields to AXPs. All these show that continuous increase of 
magnetic
field strength for single pulsars do not lead to AXP 
properties
continuously. There is the necessity to have another parameter which 
distinguish 
SGRs and AXPs from PSRs, apart from the magnetic field strength. These 
parameter may be the mass of the NS-magnetars.

For understanding the physical nature of AXPs and SGRs it is very important 
to know their position on the P$-$\.{P} diagram and whether they have 
connection 
with SNRs or not. These factors limit the age of AXPs and SGRs and give the 
foundation for discussion on their evolutionary tracks on the P$-$\.{P} 
diagram. 

SGRs do not show $directly$ observable influence on the surrounding matter 
which may support such large value of \.{P}. 
Therefore they must be 
magnetars and have age not more than AXPs if we do not consider accretion. 
Negative result of SNR 
search around these objects do not necessarily show that their ages are more 
than 
(2-5)$\times 10^4$ years because we cannot observe many SNRs in strong 
star formation regions. 
It is necessary to take into consideration that for PSRs J1124-5916, 
J1614-5047, J1617-5055 and J1301-6305 which have ages $<$10$^4$ years 
and have distances $\sim$7-8 kpc, no SNR connected with them have been found 
yet.

As all the PSRs which have genetic connections with SNRs have 
periods considerably smaller than 1 sec, it is
natural that AXPs and SGRs at birth also have periods less than 1 sec. For 
them 
to reach the current positions in the P-\.{P} diagram, with whatever 
process (accretion, magnetar, or any other complicated model), they must 
have much 
larger \.{P} values in earlier times. If that was not the case they 
should be much older and they would not have connections with SNRs.

AXP 1E2259+586 and AXP 1E1841-045 which have definite connections with  
SNRs have characteristic ages of 200 and 4.7 kyr, respectively (Table 
8). The SNRs
G109.1-1.0 and G27.4+0.0 connected with these AXPs have ages 3-10 kyr  
[44,45,18,5]
and 2 kyr
[46,47,18,5],
respectively. For the SNR G27.4+0.0 and AXP1E1841-0.45 pair, SNR age 
and characteristic age
difference is in the error limits. For the pair SNR G109.1-1.0 and AXP
1E2259+586, this difference is not less than a factor of 20 and today we 
do not have observations showing a sign of magnetic decay (except the 
period noise), i.e. there are 
no bursts observed. However, there might be unobservable less energetic 
frequent bursts.

In the
magnetar model, characteristic time of magnetic decay of the NS must be  
smaller than $\tau$ in the region on the P-\.{P} diagram where AXPs and 
SGRs are
located today (see Figure 8). To equate the ages of AXP1E2259+586 and SNR 
G109.1-1.0,  it is necessary to have strong decay of magnetic fields and 
the initial magnetic field of this AXP must not be smaller than that of 
SGRs'.
As SGRs and AXPs are located on the P$-$\.{P} diagram practically on 
a vertical line and there is no AXPs/SGRs with larger periods, we must talk 
about 
their evolution with very large value of braking index, n which might be 
caused by magnetic decay. 
It is strange not only we do not know such a decay among the PSRs with 
large magnetic fields, but also the absence of bursts which we may 
observe in any radiation band [48].

DRQNS RXJ0720.4-3125, RXJ0420.0-5022 and RXJ0806.4-4123 are the 
continuations of evolutionary tracks of SGRs and AXPs on the P-\.{P} 
diagram with large value of n. 
These two sources do not have measured values of \.{P} (see Table 11). But 
these sources are located very close, maybe d$<$ 0.8 kpc and 
located 
in Puppis-Vela direction at high enough latitude, b. Therefore 
absence of the SNR which is connected with these sources puts lower limit 
for their age; approximately 3 10$^4$ years. From the known values of P 
and t$>$3 10$^4$ years, it follows that \.{P} for RXJ0806.4-4123 and 
RXJ0420.0-5020 respectively are smaller than 6 10$^{-12}$ s/s and 1.2 
10$^{-11}$ s/s.
These sources have 
kT$\sim$0.06-0.09 keV and have thermal radiation luminosity 
$\sim$10$^{32}$ erg/s in 0.1-2.4 keV band (see Table 11). 
Therefore their ages must be small; in this case they cannot be result of 
evolution of ordinary radio PSRs. If in reality SGRs and AXPs have evolution 
with very 
large braking index then these thermal sources may be the final stage of 
evolution of SGRs and have ages $\sim$10$^5$ years. Such value of age do 
not contradict the cooling times of NSs [49,50].
But if these sources appear in result of evolution of massive PSRs with 
magnetic 
field about 10$^{13}$ Gauss then their ages must be $>$10$^6$ years and 
their luminosity $<$10$^{30}$ erg/s. But observed temperatures and 
luminosities of DRQNSs with large periods may be result of evolution of 
low mass NS with large magnetic fields.

On the other hand, number density of DRQNSs with large P is high; 4 objects 
at d$<$0.8 kpc (Table 11). Therefore we can say that there is a 
large number of NSs born with B$>$10$^{13}$ Gauss. 
But this 
and also the value of \.{P} (which contain large error) of RBS 1223 do 
not yet justify the magnetar model with NS mass about 
1M$_\odot$. 

\section{Discussion and Conclusion}
1. The absolute deviations of period values from the constant \.{P} line 
in period histories are larger for 
SGR1900+14, SGR1806-20 and AXP1048.1-5937 which have the largest values of 
\.{E} in comparison to all other AXPs. 
In the frame of the magnetar model, these objects 
must also have large values of temperature and persistent luminosities 
since they have higher magnetic field decay. However, these sources do 
not have the largest temperature and luminosity among AXPs on the average. 
This is hard to 
understand in the frame of the magnetar model. Moreover, we expect to 
see the same behavior of unobserved bursts and noise  
from AXPs 1E1841-045 and RXJ170849-4009 similar to AXP 1048.1-5937, as the 
number of period data 
increase. If burst and noise characteristics are not observed from these 
AXPs, then this would be a handicap for the magnetar model. 

2. We estimated the lower limit of \.{P} value of AXP J1845.0-0300 to be 
about 1.3 
10$^{-12}$ s/s using the SNR age as a constraint. The existence of magnetic 
decay decreases the upper limit of \.{P}. We conclude that \.{P} should 
be in the range between (1.3-2) 10$^{-12}$ s/s.

3. The large increase in the P and \.{P} values of SGR 1900+14 and SGR 
1806-20 after the burst activity periods might be caused by plasma 
ejection and 
interaction with the surrounding matter. The time-scale 
of increase of P which is about 1-3 years for these two objects, shows that 
the 
process increasing the period finishes its effect when the ejected 
plasma reaches less than 1 pc from the NS. The high P values 
of these SGRs are expected to fall to lower values in the future.

4. SGRs and AXPs are not sources having strong PWN in spite of the 
existence of bursts for SGRs and possible burst activity for AXPs. It is 
shown that PWN appears only 
if pulsar have small ages ($<$10$^4$ years) and simultaneously very large 
value of \.{E} (which is proportional to voltage). Nearly all PSRs with 
$\tau < 1.6\times10^4$ years up to 
distance 5 kpc were found to be associated with SNRs. All 21 PSRs which have 
connection with 
SNRs, have values of \.{E}$>$3$\times10^{34}$ erg/s. But PWN have only 
been observed from 8 
PSRs which have $\tau < 3\times 10^4$ years and \.{E}$> 4\times10^{35}$ 
erg/s. On the other hand, near the SGR1900+14, SGR1806-20, AXP1048.1-5937 
and AXPJ170849-4009 with characteristic ages $<$10$^4$ years and 
distances d$<$10-15 kpc even no S-type SNRs are found. Evidently, it is 
not easy to find SNRs at such distances in the galactic center 
directions. But absence of PWN may be natural because for these 
objects \.{E}$<$4$\times 10^{34}$ erg/s. 

5. Magnetar model considered SGRs and AXPs as one group of objects, so that 
AXP 1E2259+586 must have lost burst activity and underwent 
considerable magnetic decay. If we equate the characteristic age of this 
AXP to the age of 
SNR CTB 109, it is necessary for the initial magnetic field not to be 
smaller than the magnetic field of SGRs. From this fact and the location of 
SGRs and AXPs on 
the P-\.{P} diagram (Figure 8) roughly on a hypothetical vertical line 
(P=const.) we may 
talk about strong magnetic field decay and large value of braking index for 
AXP/SGR group. It is also necessary to note that if source of burst 
energy is the magnetic 
field decay and because several DRQNSs are
in the continuation of the evolutionary 
tracks of AXPs/SGRs, then the decay of magnetic field of AXPs and SGRs must 
continue further. 
If the DRQNSs with large periods, are the 
continuation of AXPs/SGRs, then we expect to observe noise from these 
pulsars higher than the noise of radio PSRs.
Also, we expect to observe high noise from newly found PSRs which have 
magnetic fields closer to AXP 1E2259+586.

6. In Thompson \& Duncan [4] model of magnetar, the order of magnitude 
of the magnetic field is estimated from P-\.{P} values and also from the 
strength of the giant bursts. If these sources are involved in several 
bursts before, then they need to have several times higher magnetic field 
strength at birth than the current magnetic field strength or their magnetic 
field should be fed continuously by dynamo action. 
It is a possibility that the NSs with SGR and AXP properties might 
have lower 
mass (so larger in radius) and B$\sim$10$^{13}$-10$^{14}$ G) 
to have burst activity and such noise. This makes 
the drift of monopoles much  easier, characteristic time for turbulence 
increases and consequently dynamo action works longer and magnetic field 
is fed longer. The NS with low mass perform pulsations for a long time 
until it comes to stability. This makes them cool later and increase the 
turbulence activity which feeds the dynamo.

7. Continuous increase of magnetic field for PSRs do not lead to AXP 
properties continuously. If in addition to the high magnetic fields we 
assume that the NS has lower mass, then, this causes \.{P} to 
decrease sharply when NS reaches stability. Consequently the evolution 
from SGR to AXP and then to DRQNS do not require strong magnetic decay. In 
this case, we can say that there is no fundamental difference between PSRs 
and magnetars.

$Acknowledgments$
We thank T\"{U}B\.{I}TAK, the Scientific and Technical Research
Council of Turkey, for support through TBAG-\c{C}G4. We also would like 
to thank M. \"{O}zg\"{u}r Ta\c{s}k\i n for comments on the manuscript.

 \clearpage

\begin{table}
\footnotesize
\begin{tabular}{lllll}\hline\hline
\multicolumn{1}{l}{\ Day} &
\multicolumn{1}{l}{\ Period (s)} &
\multicolumn{1}{l}{\ Uncert.} &
\multicolumn{1}{l}{\ Observatory} &
\multicolumn{1}{l}{\ Reference} \\
\hline
1978-01-14 & 6.978586  &  0.000006  &  HEAO-1
&\hspace{0.5cm} 51\\
1980-06-08 & 6.97864   &  0.00003   &  Einstein  &
\hspace{0.5cm}   52\\
1981-01-24 & 6.978632  &  0.000014  &  Einstein  &
\hspace{0.5cm}   52\\
1983-10-12 & 6.978675  &  0.000010  &  Tenma     &
\hspace{0.5cm}  53\\
1984-12-01 & 6.978720  &  0.000006  &  EXOSAT     &
\hspace{0.5cm}  54\\
1984-12-01 & 6.978725  &  0.000008  &  EXOSAT     &
\hspace{0.5cm}  55\\
1987-06-22 & 6.978759  &  0.000002  &  Ginga     &
\hspace{0.5cm}  56\\ 
1989-12-15 & 6.978789  &  0.000007  & Ginga      &
\hspace{0.5cm}  57\\
1990-08-09 & 6.978795  &  0.000002  & Ginga      & 
\hspace{0.5cm}  57\\
1991-07-08 & 6.978818  &  0.000004  &  Rosat     & 
\hspace{0.5cm}  58\\ 
1992-01-07 & 6.978806  &  0.000006  &  Rosat     &
\hspace{0.5cm}  58\\ 
1992-06-24 & 6.978824  &  0.000007  &  Rosat     &
\hspace{0.5cm}   58\\
1993-02-06 & 6.978837  &  0.000006  &  Rosat     &
\hspace{0.5cm}  58\\
1993-05-30 & 6.9788465 &  0.0000056 &  ASCA      & 
\hspace{0.5cm}   58\\
1996-11-16 &  6.978914 &   0.000006 &    BeppoSAX &
\hspace{0.5cm}   59\\
1997-03-09 & 6.978919950 & 0.000000088 & RossiXTE  &
\hspace{0.5cm}  60\\ 
2000-01-11 & 6.978977  & 0.00000024 & Chandra &
\hspace{0.5cm}  61\\
\hline
\end{tabular}

{Table 1: AXP  1E2259+586} \\
\end{table}

\begin{table}
\footnotesize
\begin{tabular}{lllll}\hline\hline
\multicolumn{1}{l}{\ Day} &
\multicolumn{1}{l}{\ Period (s)} &
\multicolumn{1}{l}{\ Uncert.} &
\multicolumn{1}{l}{\ Observatory} &
\multicolumn{1}{l}{\ Reference} \\
\hline
1987-05-09 & 11.758320 & 0.000007 & Ginga
&\hspace{0.5cm}       62\\
1991-04-24 & 11.76346  & 0.000023 & Ginga
&\hspace{0.5cm}       62\\
1992-03-16 & 11.7645   & 0.0      & ROSAT
&\hspace{0.5cm}       62\\
1993-10-11 & 11.76676  & 0.000045 & ASCA
&\hspace{0.5cm}       62\\
1996-08-31 & 11.7707   & 0.000375 & RXTE
&\hspace{0.5cm}       62\\
1998-03-27 & 11.77248  & 0.000050 & ASCA
&\hspace{0.5cm}       62\\  
1999-04-08 & 11.77387  & 0.000034 & BeppoSAX
&\hspace{0.5cm}       62\\
\hline
\end{tabular}

{Table 2: AXP 1E1841-045} \\
\end{table}
\clearpage

\begin{table}
\footnotesize
\begin{tabular}{lllll}\hline\hline
\multicolumn{1}{l}{\ Day} &
\multicolumn{1}{l}{\ Period (s)} &
\multicolumn{1}{l}{\ Uncert.} &
\multicolumn{1}{l}{\ Observatory} &
\multicolumn{1}{l}{\ Reference} \\
\hline
1979-07-14 &    6.4377    & 0.001    &  Einstein
&\hspace{0.5cm} 63\\
1985-07-17 &    6.4407    & 0.0009   &  EXOSAT
&\hspace{0.5cm} 63\\
1988-04-15 &    6.44153   & 0.0002   &  Ginga
&\hspace{0.5cm} 64\\
1988-04-25 &    6.44185   & 0.00001  &  Ginga
&\hspace{0.5cm} 64\\
1988-09-28 &    6.4422    & 0.0004   &  Ginga
&\hspace{0.5cm} 64\\
1992-06-13 &    6.444868  & 0.000007 &  ROSAT
&\hspace{0.5cm} 65\\
1992-08-10 &    6.44499   & 0.000034 &  ROSAT
&\hspace{0.5cm} 65\\
1992-12-16 &    6.44532   & 0.000072 &  ROSAT
&\hspace{0.5cm} 65\\
1993-01-07 &    6.445391  & 0.000013 &  ROSAT
&\hspace{0.5cm} 65\\
1994-03-05 &    6.446645  & 0.000001 &  ASCA     &
\hspace{0.5cm} 66,67\\
1996-07-30 &    6.449769  & 0.000004 &  RossiXTE
&\hspace{0.5cm} 68\\
1997-01-25 &    6.4502054 & 0.0000001 &  RossiXTE
&\hspace{0.5cm} 69\\
1997-03-09 &    6.450198  & 0.000018  & RossiXTE
&\hspace{0.5cm} 70\\
1997-05-10 &    6.45026  &  0.000013 &  BeppoSAX
&\hspace{0.5cm} 71\\
1997-07-17 &    6.4504252 & 0.0000001 & RossiXTE &
\hspace{0.5cm} 69\\
1997-10-09 &    6.450484  & 0.000018  & RossiXTE &
\hspace{0.5cm}   70\\
1998-07-27 &    6.450815  & 0.000002  & ASCA    &
\hspace{0.5cm}   67\\
1999-01-23 &    6.45129397 & 0.00000005 & RossiXTE
&\hspace{0.5cm}  69\\
2000-04-11 &    6.4520766 & 0.0000005 & RossiXTE  &
\hspace{0.5cm}   69\\
\hline
\end{tabular}

{Table 3: AXP 1E1048.1-5937} \\
\end{table}

\begin{table}
\footnotesize
\begin{tabular}{lllll}\hline\hline
\multicolumn{1}{l}{\ Day} &
\multicolumn{1}{l}{\ Period (s)} &
\multicolumn{1}{l}{\ Uncert.} &
\multicolumn{1}{l}{\ Observatory} &
\multicolumn{1}{l}{\ Reference} \\  
\hline
1979-03-01 &   8.68707 &  0.00012 & Einstein &\hspace{0.4cm} 72 \\
1979-09-03 &   8.68736 &  0.0007  & Einstein &\hspace{0.4cm} 72 \\
1984-08-27 &   8.68723 &  0.00004 & EXOSAT   &\hspace{0.4cm} 73 \\
1993-02-12 &   8.68784 &  0.00004 & ROSAT    &\hspace{0.4cm} 74 \\
1994-09-18 &   8.68791 &  0.00015 & ASCA     &\hspace{0.4cm} 72 \\
1994-09-19 &   8.687873 &  0.000034 & ASCA   &\hspace{0.4cm} 67 \\
1996-03-25 &   8.6881  &  0.0002  & RXTE     &\hspace{0.4cm} 75 \\
1996-03-28 &   8.688068 &  0.000002 & RXTE   &\hspace{0.4cm} 15 \\
1997-01-03 &   8.68804 &  0.00007 & BeppoSAX &\hspace{0.4cm} 15 \\
1997-08-09 &   8.6882  &  0.0002  & BeppoSAX &\hspace{0.4cm} 15 \\
1998-02-03 &   8.6883  &  0.0001  & BeppoSAX &\hspace{0.4cm} 15 \\
1998-08-22 &   8.688267 &  0.000024 & ASCA   &\hspace{0.4cm} 67 \\
\hline
\end{tabular}

{Table 4: AXP 4U0142+61}
\end{table}
\clearpage

\begin{table}
\footnotesize
\begin{tabular}{lllll}\hline\hline
\multicolumn{1}{l}{\ Day} &
\multicolumn{1}{l}{\ Period (s)} &
\multicolumn{1}{l}{\ Uncert.} &
\multicolumn{1}{l}{\ Observatory} &
\multicolumn{1}{l}{\ Reference} \\
\hline
1996-09-03 & 10.99758    & 0.00005   & ASCA  & 14 \\
1997-03-25 & 10.99802    & 0.00005   & Rosat & 76 \\
1999-02-07 & 10.99905117 & 0.0000006 & RXTE  & 77 \\
1999-02-07 & 10.99903546 & 0.0000006 & RXTE  & 75 \\
\hline
\end{tabular}

{Table 5: AXP 1RX J1708-4009}
\end{table}

\clearpage
\begin{table}
\footnotesize
\begin{tabular}{lllll}\hline\hline
\multicolumn{1}{l}{\ Day} &
\multicolumn{1}{l}{\ Period (s)} &
\multicolumn{1}{l}{\ Uncert.} &
\multicolumn{1}{l}{\ Observatory} &
\multicolumn{1}{l}{\ Reference} \\
\hline
1996-09-10 & 5.1558157 &  0.0000003 &  RXTE    &      22\\
1996-09-12 & 5.1558199 &  0.0000029 &  RXTE    &      18\\
1997-05-12 & 5.157190  &  0.000007  &  BeppoSAX &     79\\
1998-05-01 & 5.1589715 &  0.0000008 &  ASCA   &       80\\
1998-06-05 & 5.15917011 &  0.00000055 &  RXTE  &        22\\
1998-06-06 & 5.159142 &   0.000003  &  RXTE    &      81\\
1998-08-31 & 5.160199 &   0.000002  &  RXTE    &      81\\
1998-09-01 & 5.160197 &    0.000001  &  RXTE  &        82\\
1998-09-13 & 5.16026572 &  0.00000012 &  RXTE  &        22\\
1998-09-15 & 5.160262  &  0.000011  &  BeppoSAX &     79\\
1998-09-16 & 5.160295  &  0.000003  &  ASCA    &      83\\
1998-12-12 & 5.16129785 &  0.00000008 &  BSA   &        84\\
1999-01-03 & 5.160934 &   0.000056  &  RXTE    &      22\\
1999-03-21 & 5.16145  &   0.00018  &   RXTE    &      22\\
1999-03-30 & 5.16156  &   0.00011  &   RXTE    &      22\\
2001-04-21 & 5.17274   &  0.00022  &   RXTE    &      85\\
2001-04-22 & 5.172908 &   0.000040  &   Chandra &       85\\
2001-04-30 & 5.172947 &   0.000065 &   Chandra  &     85\\
2001-04-30 & 5.17321  &   0.00021  &   RXTE     &     85\\
\hline
\end{tabular}

{Table 6: SGR1900+14}
\end{table}

\clearpage

\begin{table}
\footnotesize
\begin{tabular}{lllll}\hline\hline
\multicolumn{1}{l}{\ Day} &
\multicolumn{1}{l}{\ Period (s)} &
\multicolumn{1}{l}{\ Uncert.} &
\multicolumn{1}{l}{\ Observatory} &
\multicolumn{1}{l}{\ Reference} \\
\hline
1993-10-15 & 7.46851 & 0.00025 & ASCA & 86\\        
1995-10-16 & 7.4738 & 0.001 & ASCA & 86\\           
1996-11-05 & 7.476551 & 0.000003 & RXTE & 86\\     
1996-11-11 & 7.4765534 & 0.0000012 & RXTE & 29\\    
1998-10-16 & 7.48175 & 0.00016 & BeppoSAX & 87\\     
1999-03-21 & 7.48271 & 0.00003 & BeppoSAX & 87\\   
1999-05-02 & 7.48315368 & 0.00000029 & RXTE & 29\\
2000-08-16 & 7.4925  & 0.0002 & CHANDRA & 31\\
\hline
\end{tabular}

{Table 7: SGR1806-20}
\end{table}

\clearpage
\newpage

\begin{figure*}
\centerline{\psfig{file=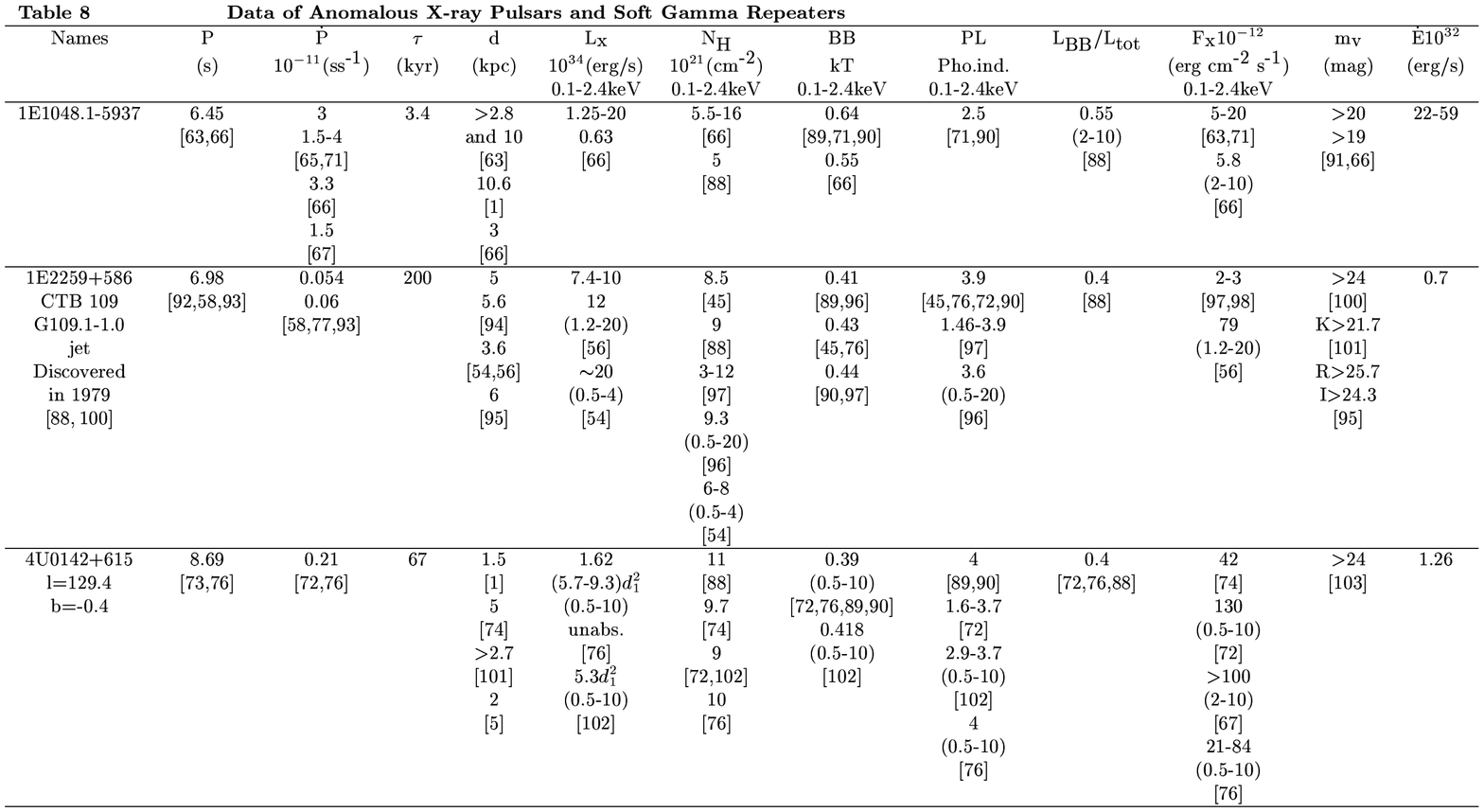,width=18cm,height=20cm,angle=-90}}
\end{figure*}

\begin{figure*}
\centerline{\psfig{file=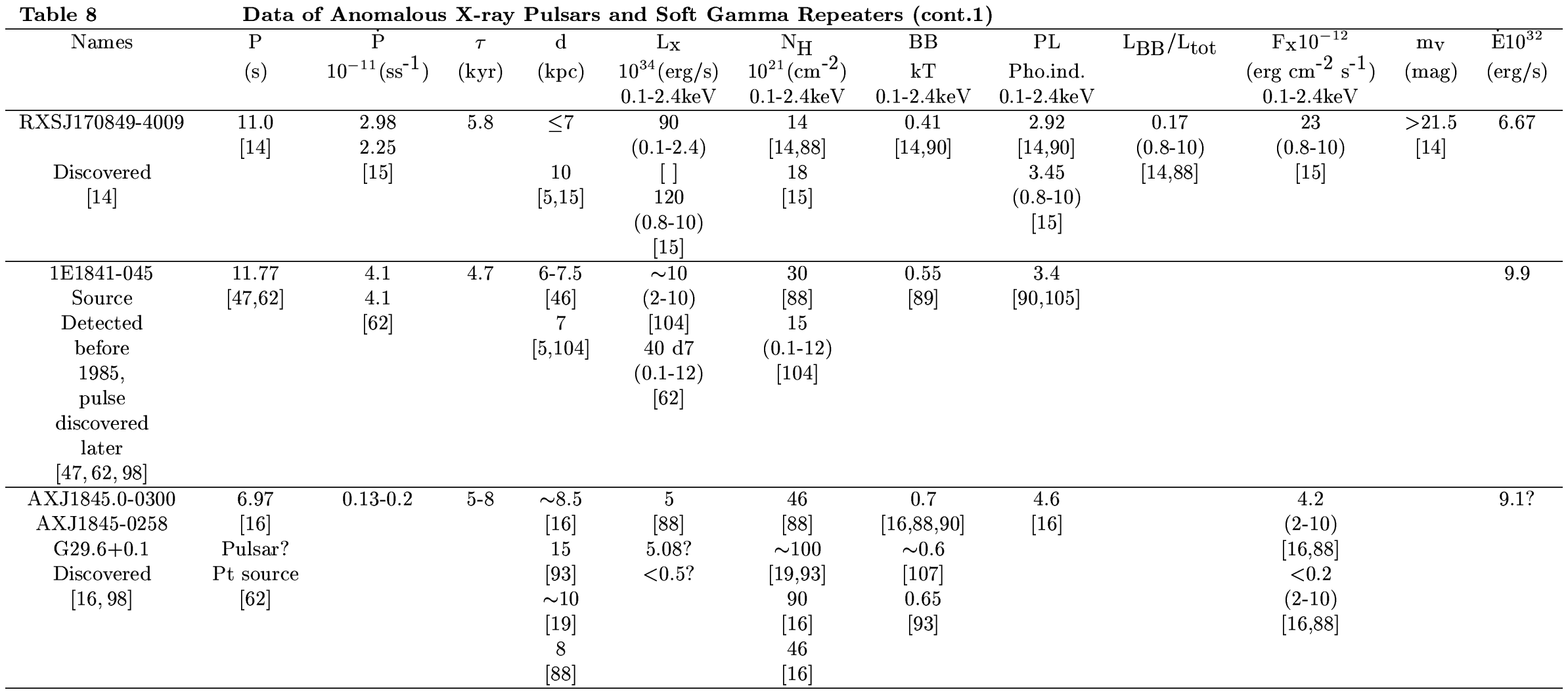,width=18cm,height=20cm,angle=-90}}
\end{figure*}

\begin{figure*}
\centerline{\psfig{file=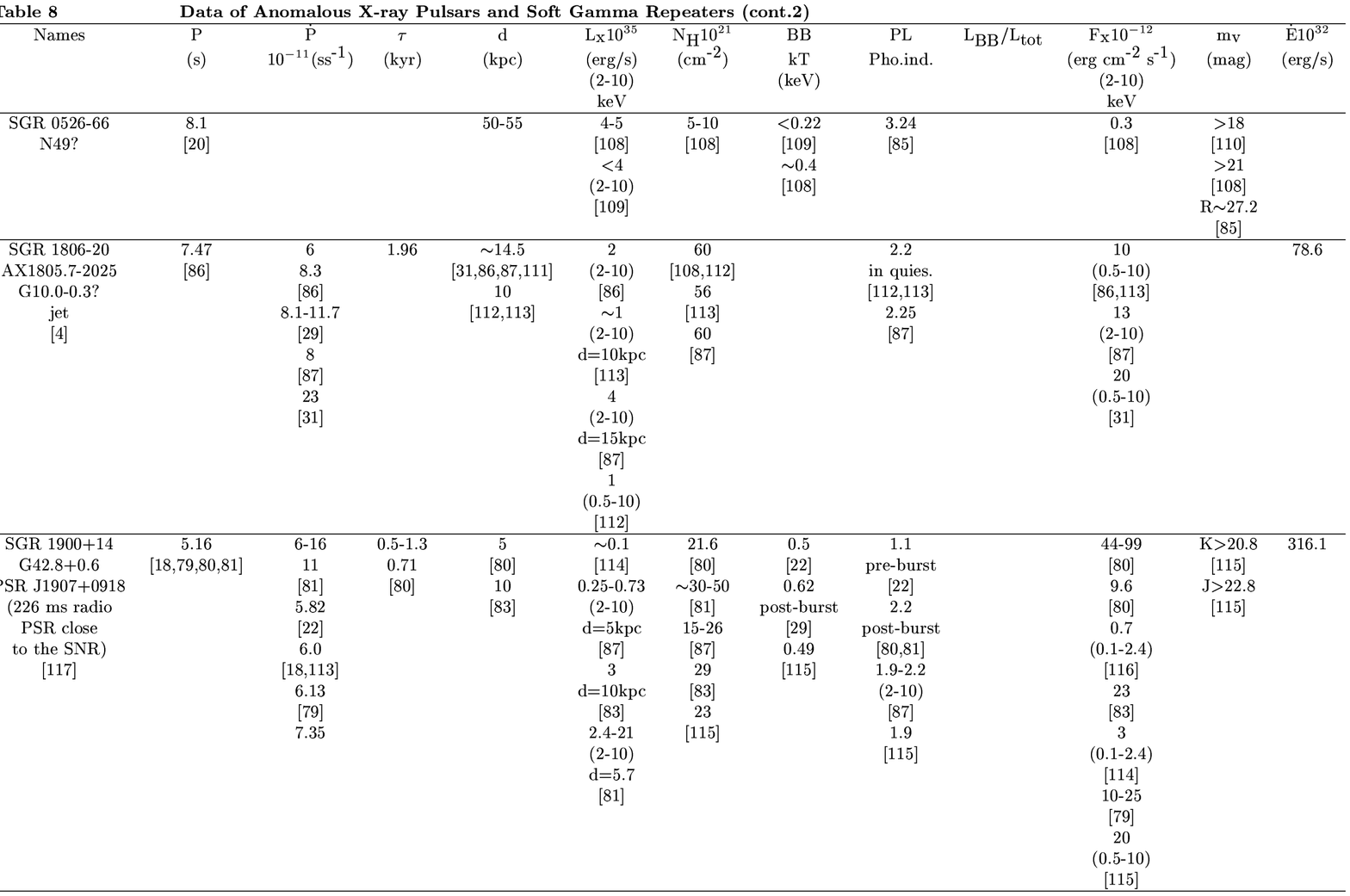,width=18cm,height=20cm,angle=-90}}
\end{figure*}

\begin{figure*}
\centerline{\psfig{file=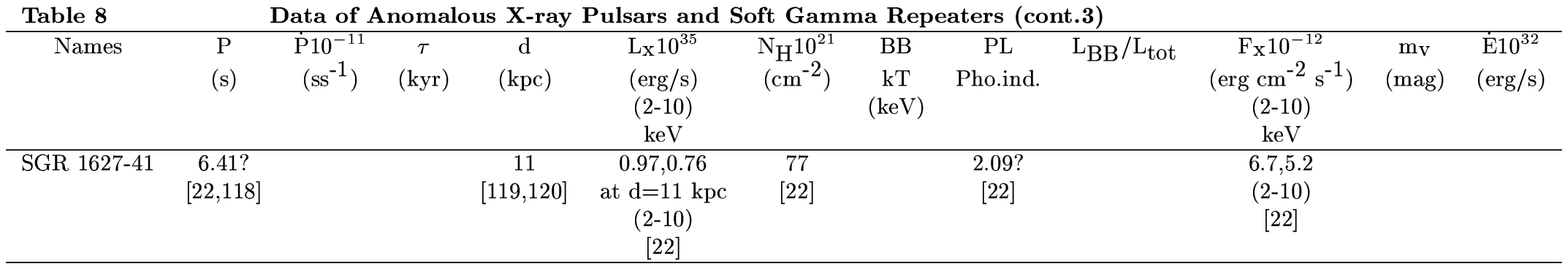,width=18cm,height=17cm,angle=-90}}
\end{figure*}

\clearpage
\newpage

\begin{table}[h]     
\footnotesize
\begin{tabular}{llll}\hline\hline
\multicolumn{1}{l}{\ Name} &
\multicolumn{1}{l}{\ Burst Energy (erg/s)} &
\multicolumn{1}{l}{\ Intensity (erg/cm$^2$s)} &
\multicolumn{1}{l}{\ Reference} \\
\hline
SGR 1806-20              &                   &                 		& \\
1993-10-07 -- 1993-12-12 & 10$^{39-41}$      &	(1.4-430) 10$^{-8}$	& 32\\ 
1995-09-23 -- 1995-10-27 &	             &	                        & \\
1996-10-19 -- 1998-02-06 &                   &  	                & \\
1998-07-06 -- 1999-07-08 &                   &                      	& \\
2000-08-07               &                   &                          & 121\\
 \hline
SGR 1900+14              &                   &  	        	& \\
1998-05-26--30           &                   &	1.6$10^{-5}$	       &24\\
1998-08-27               &$\sim$5 $10^{42}$    &$>$2.1 $10^{-4}$ &23\\ 
1998-06-2 -- 1998-12-21  &                   &$\sim$10$^{-6}$ 	&24\\ 
2001-04-18               &                   &1.1$10^{-5}$&25\\   
2001-07-2                &$\sim$7 $10^{40}$ &1.4$10^{-6}$  &26\\ 
\hline
\end{tabular}

{Table 9. Burst Data of SGRs which have spin period histories}
\end{table}

\begin{table}
\footnotesize
\rotate{
\begin{tabular}{llllllllllll}\hline\hline
\multicolumn{1}{l}{\ PSR J} &
\multicolumn{1}{l}{\ Pulsed} &
\multicolumn{1}{l}{\ log L$_{400}$} &
\multicolumn{1}{l}{\ log L$_{1400}$} &
\multicolumn{1}{l}{\ log L$_x$} &
\multicolumn{1}{l}{\ \.{E}} &
\multicolumn{1}{l}{\ SNR} &
\multicolumn{1}{l}{\ Type} &
\multicolumn{1}{l}{\ L$_{tot}$ 10$^{35}$} &
\multicolumn{1}{l}{\ d} &
\multicolumn{1}{l}{\ D} &
\multicolumn{1}{l}{\ n} \\
    & Radiation & & & & (erg/s) & & & (erg/s) & (kpc) & (pc) & \\
\hline
0534+2200 & R, O, X, $\gamma$ & 3.41 & 1.75 & 36.0 & 38.6 & Crab & F     & 
1.8  & 2    & 3.5 & 2.5 \\
0540-6919 & R, O, X, $\gamma$ & 3.24 &      & 36.2 & 38.2 & N157B& F     & 
3.5  & 50   & 24  & 2.2 \\
0835-4510 & R, O, $\gamma$    & 3.1  & 2.31 & 32.6 & 36.8 & Vela & C     & 
0.15 & 0.45 & 32    & 1.4 \\
1119-6127 & R                 &      & 1.70 &      & 36.4 & G292.2-05& S &
0.15 & 7.5  & 16  & 2.9 \\
1513-5908 & R, O, X, $\gamma$ & 1.55 & 1.25 & 34.3 & 37.2 & RCW 89   & C &
0.46 & 4.2  & 42    & 2.84 \\
\end{tabular}
}
{Table 10. Data taken from [122]}
\end{table}

\clearpage
\newpage

\begin{figure*}
\centerline{\psfig{file=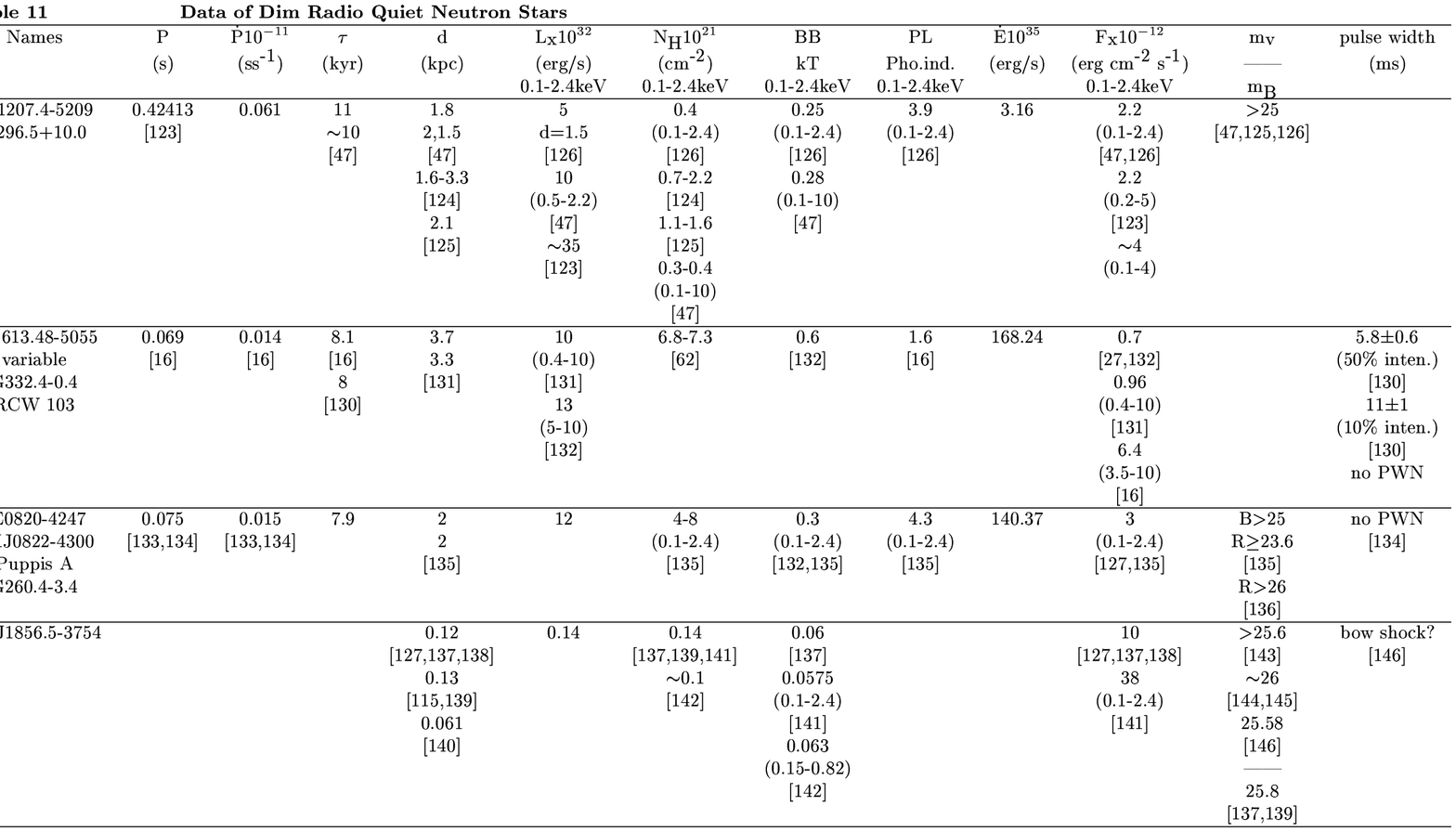,width=18cm,height=17cm,angle=-90}}
\end{figure*}

\begin{figure*}
\centerline{\psfig{file=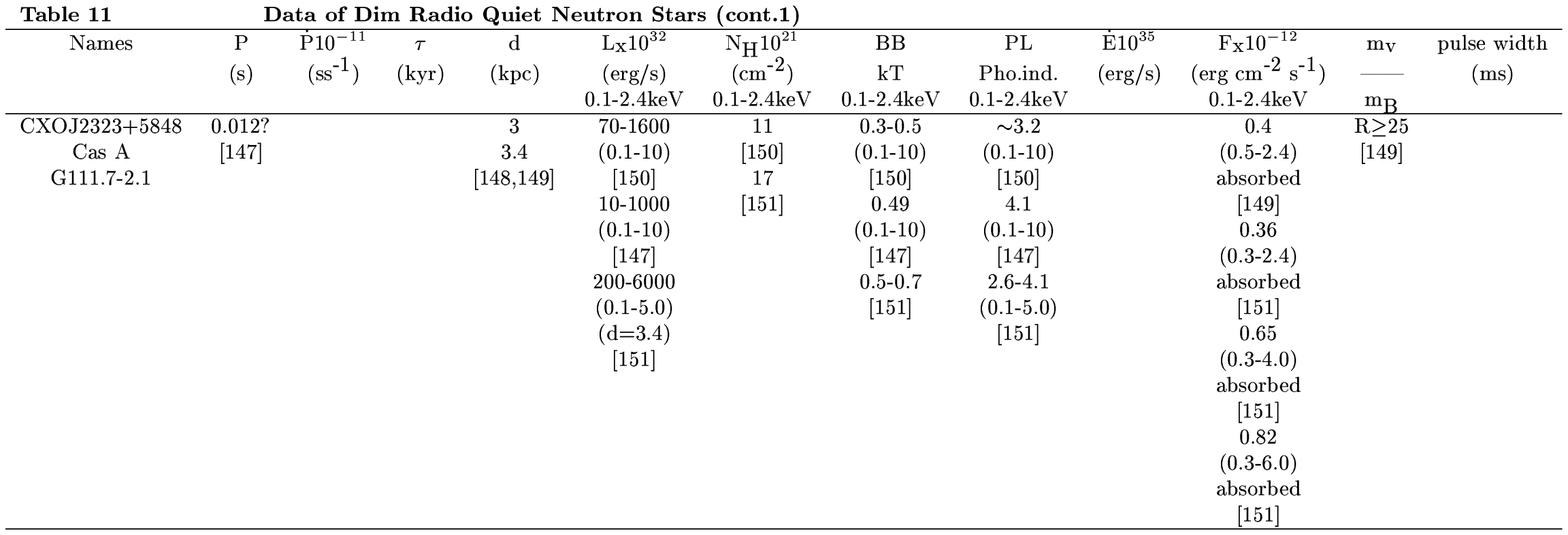,width=18cm,height=17cm,angle=-90}}
\end{figure*}

\begin{figure*}
\centerline{\psfig{file=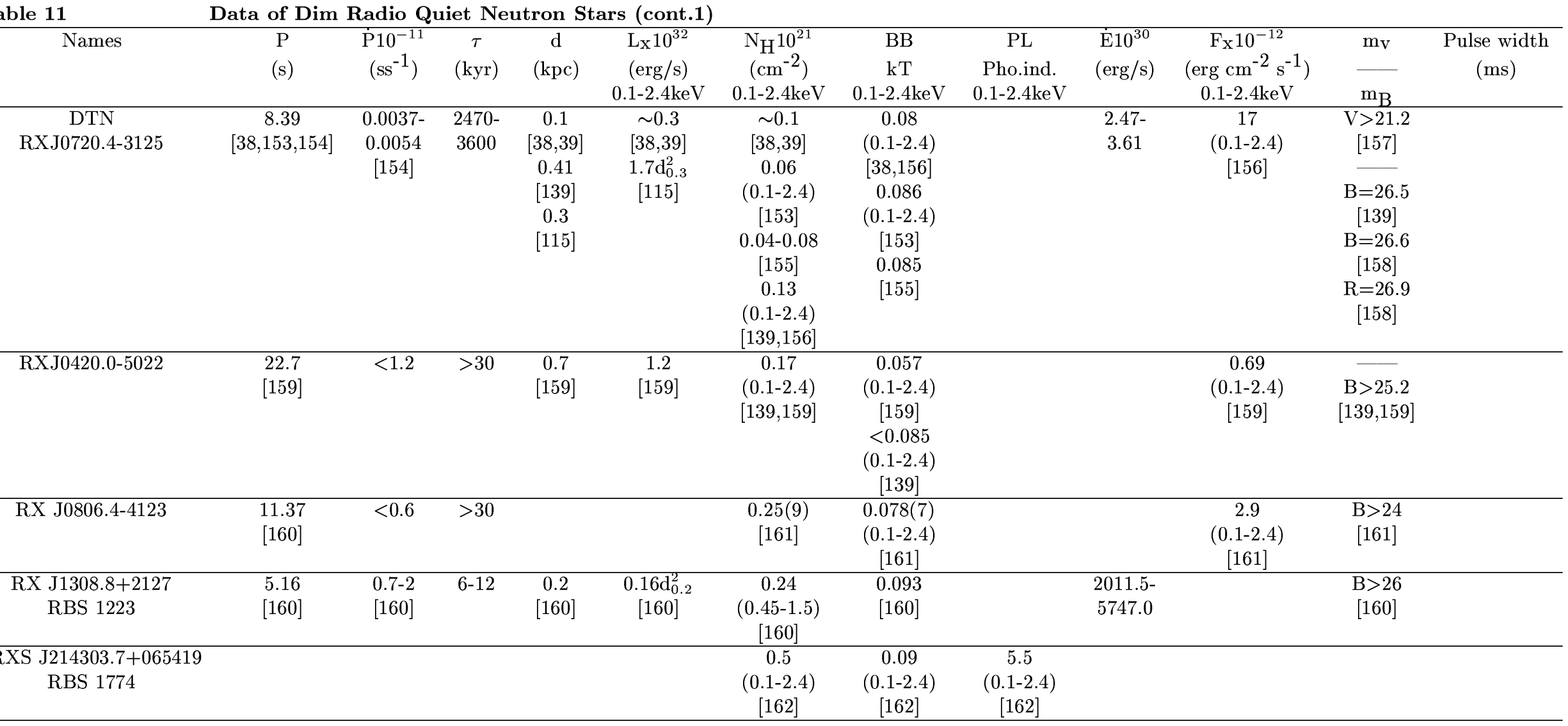,width=18cm,height=17cm,angle=-90}}
\end{figure*}

\begin{figure*}
\centerline{\psfig{file=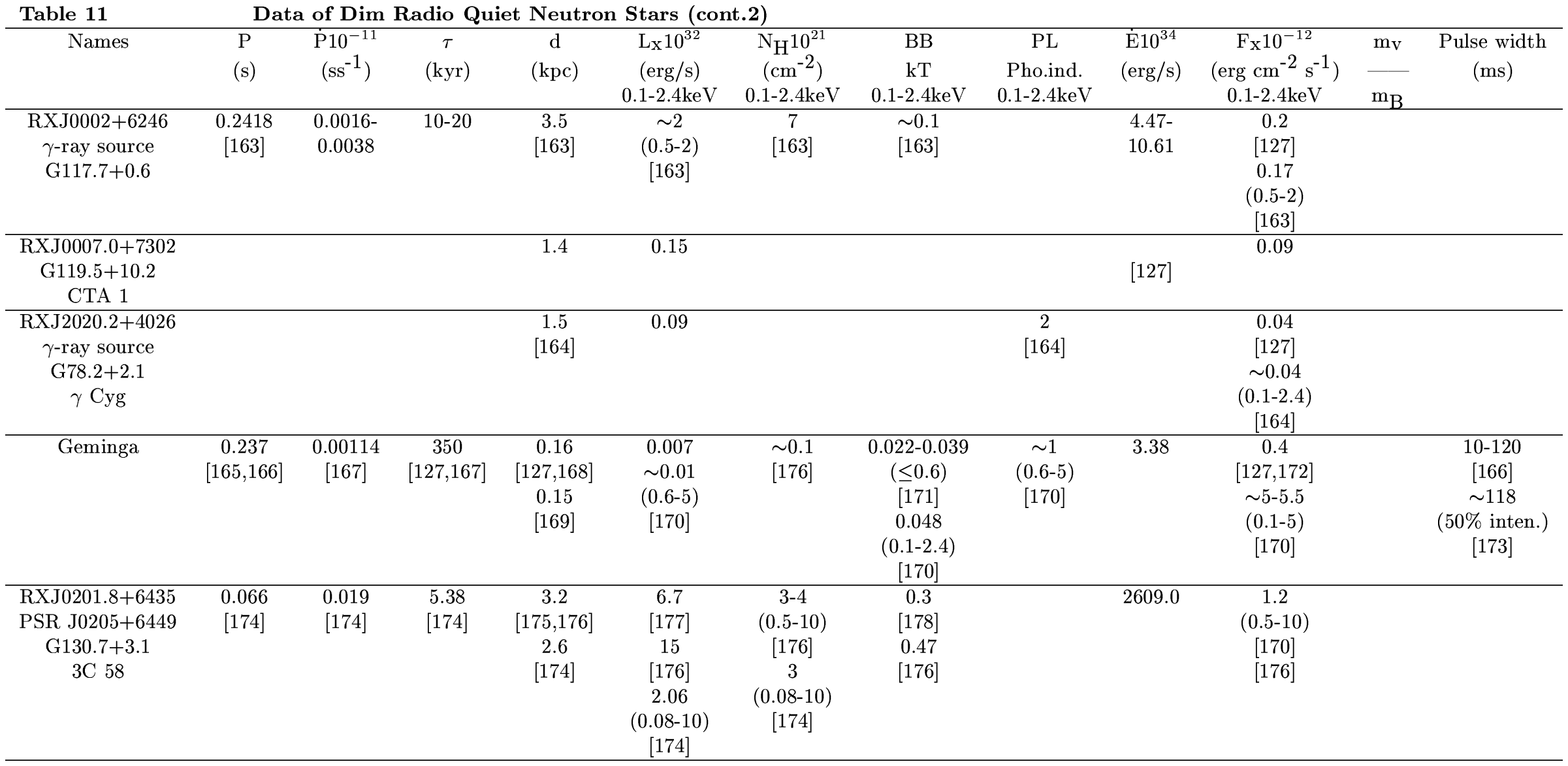,width=18cm,height=17cm,angle=-90}}
\end{figure*}

\clearpage

\begin{table}
\footnotesize
\begin{tabular}{llllll}\hline\hline
\multicolumn{1}{l}{\ PSR J} &
\multicolumn{1}{l}{\ $\tau$ (kyr)} &
\multicolumn{1}{l}{\ Total L$_x$ (erg/s)} &
\multicolumn{1}{l}{\ Pulsed L$_x$ (erg/s)} &
\multicolumn{1}{l}{\ \.{E} (erg/s)} &
\multicolumn{1}{l}{\ L$_x$/\.{E}} \\
\hline
0534-2200&1.3&1 10$^{36}$&1 10$^{36}$&4 10$^{38}$&2.5 10$^{-3}$\\
0540-6919&1.7&1.6 10$^{36}$&&1.6 10$^{38}$&1 10$^{-2}$\\
1513-5908&1.5&2 10$^{34}$&2 10$^{34}$&1.6 10$^{37}$&1.25 10$^{-3}$\\
0835-4510&11.2&4 10$^{32}$&4 10$^{31}$&6.3 10$^{36}$&6.3 10$^{-5}$\\
1952+3252&107.1&1.6 10$^{33}$&6.3 10$^{32}$&4 10$^{36}$&4 10$^{-4}$\\
1709-4428&17.4&1.6 10$^{32}$&&3.2 10$^{36}$&5 10$^{-5}$\\
2337+6151&40.7&1.2 10$^{33}$&&6.3 10$^{34}$&2 10$^{-2}$\\
0659+1414&109.6&6.3 10$^{32}$&1 10$^{32}$&4 10$^{34}$&0.016\\
1803-2137&15.8&1 10$^{33}$&&2 10$^{36}$&0.5 10$^{-3}$\\
2229+6114&10.0&5 10$^{33}$&&2 10$^{37}$&2.5 10$^{-4}$\\
0953+0755&1.7 10$^{4}$&7.9 10$^{29}$&&5 10$^{32}$&1.58 10$^{-3}$ \\
0826+2637&4.9 10$^{3}$&1 10$^{30}$&&4 10$^{32}$&2.5 10$^{-3}$\\
\hline
\end{tabular}
\\
{Table 12. Data taken from [180]}
\end{table}

\clearpage
\newpage

\begin{table}     
\footnotesize
\begin{tabular}{llll}\hline\hline
\multicolumn{1}{l}{\ Name} &
\multicolumn{1}{l}{\ Period (s)} &
\multicolumn{1}{l}{\ \.{P} (s/s)} \\
\hline
0115+634 & 3.61 &   5.9E-11 & 
\\
AXP 0142+614 & 8.687 &   2.2E-12 & 
\\
0352+309 & 836.8 &  4.2E-9 & 
\\
0532-664 & 13.5  &  6.1E-11 & 
\\
AXP 1048.1-5937 & 6.45 &   2.5E-11 & 
\\
1119-603 & 4.81 &   3.6E-11 & 
\\
1538-522 & 529  &   3.9E-9 & 
\\
1627-673 & 7.67 &   5.1E-11 & 
\\
1728-247 & 115   &  5.5E-8 & 
\\
1907+097 & 440.34 &  6.9E-9 & 
\\
2030+375 & 41.7  &  1.16E-8 & 
\\
2103.5+4545 & 358.6 &  3.2E-8 & 
\\
2138+568 & 66.25 &  1.82E-10 & 
\\
\hline
\end{tabular}

{Table 13. Data taken from [180] except AXPs which have references in Table 
3 and 4}
\end{table}
\clearpage
\newpage

%

\clearpage
\newpage

\begin{figure*}
\centerline{\psfig{file=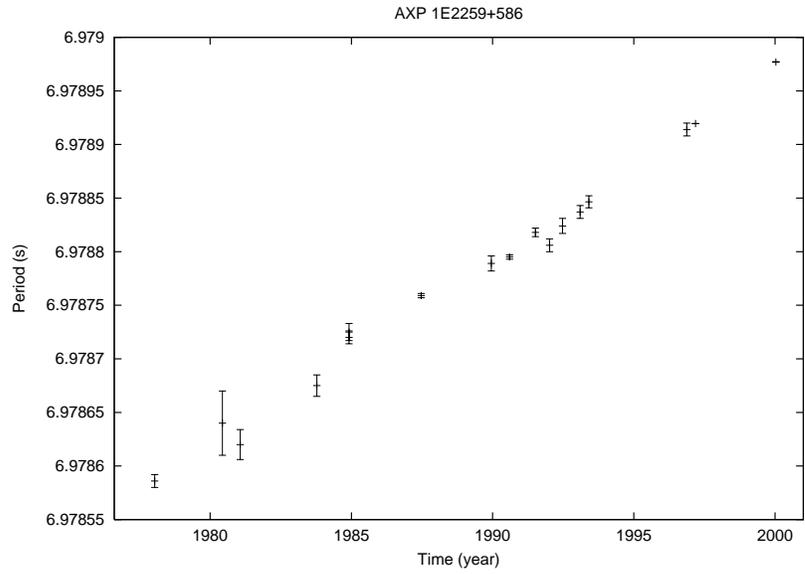,width=11cm,angle=-90}}
\caption{Period evolution of AXP 1E2259+586}
\end{figure*}

\begin{figure*}
\centerline{\psfig{file=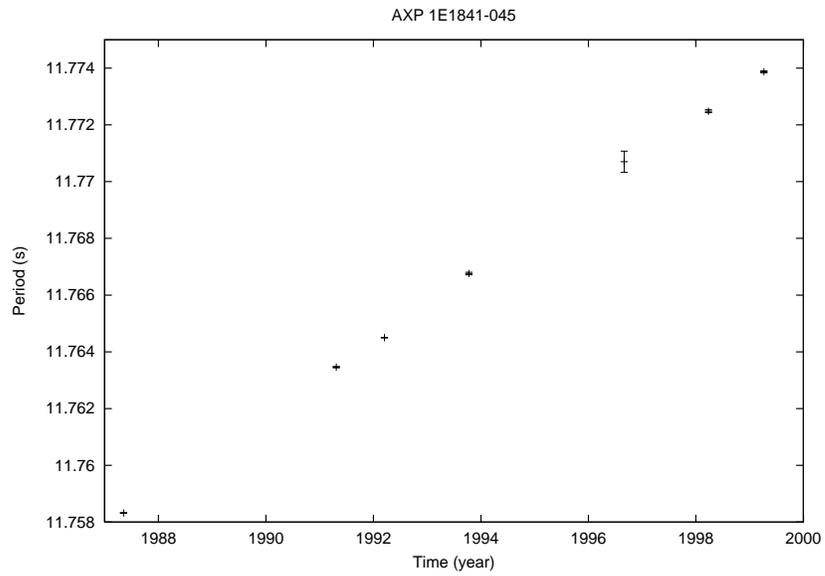,width=11cm,angle=-90}}
\caption{Period evolution of AXP 1E1841-045}
\end{figure*}

\begin{figure*}
\centerline{\psfig{file=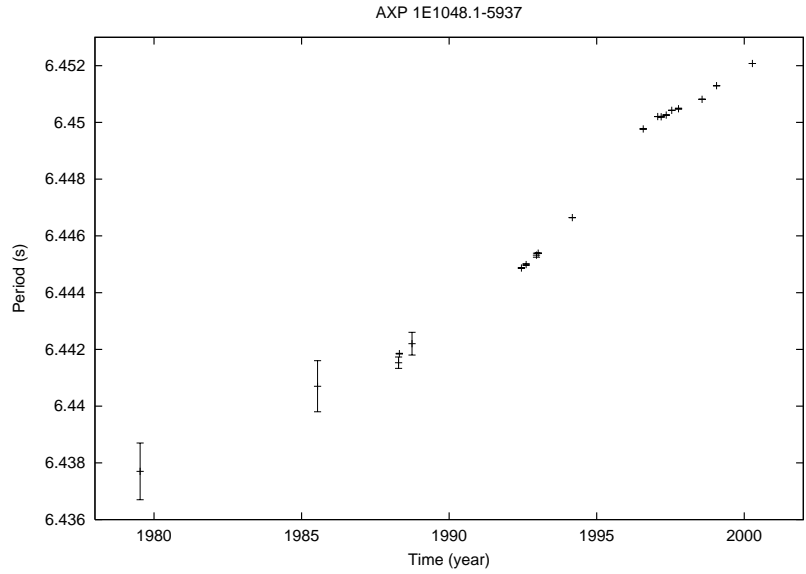,width=11cm,angle=-90}}
\caption{Period evolution of AXP 1E1048.1-5937}
\end{figure*}

\begin{figure*}
\centerline{\psfig{file=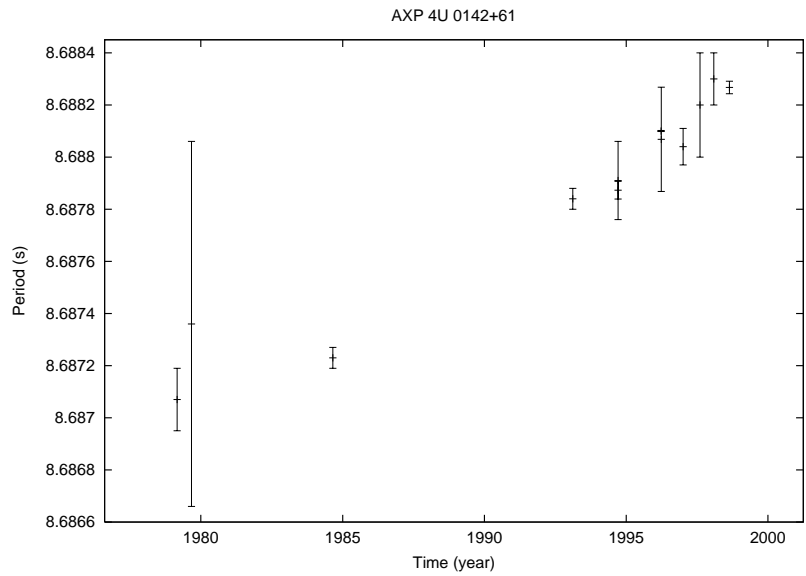,width=11cm,angle=-90}}
\caption{Period evolution of AXP 4U 0142+61}
\end{figure*}

\begin{figure*}
\centerline{\psfig{file=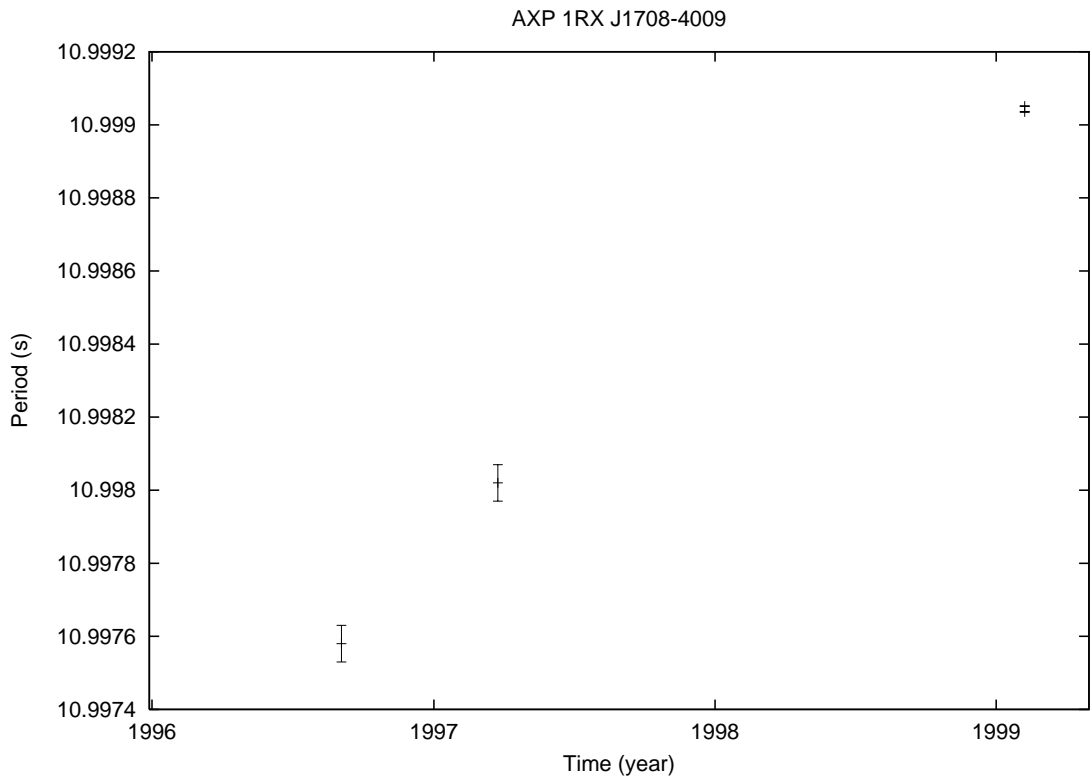,width=15cm,angle=-90}}
\caption{Period evolution of AXP 1RXJ1708-4009}
\end{figure*}

\begin{figure*}
\centerline{\psfig{file=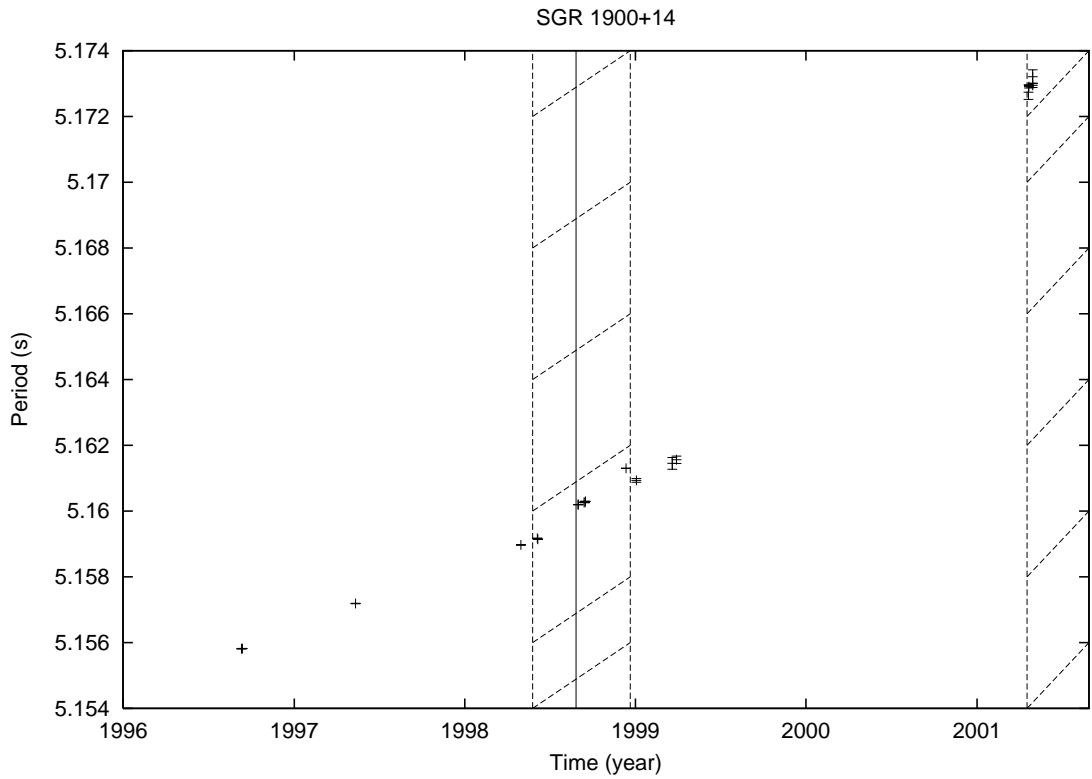,width=15cm,angle=-90}}
\caption{Period evolution of SGR 1900+14. The line constructed by +'s
represent the time of the superburst of 1998 August 27. The region
between the dashed lines near year 1998 and 1999 represent the active
period starting on 1998 May 26 and ending on 1999 February 3. The dashed
line near 2001 represent the start of another active period which started
on 2001 April 18.}

\end{figure*}

\begin{figure*}
\centerline{\psfig{file=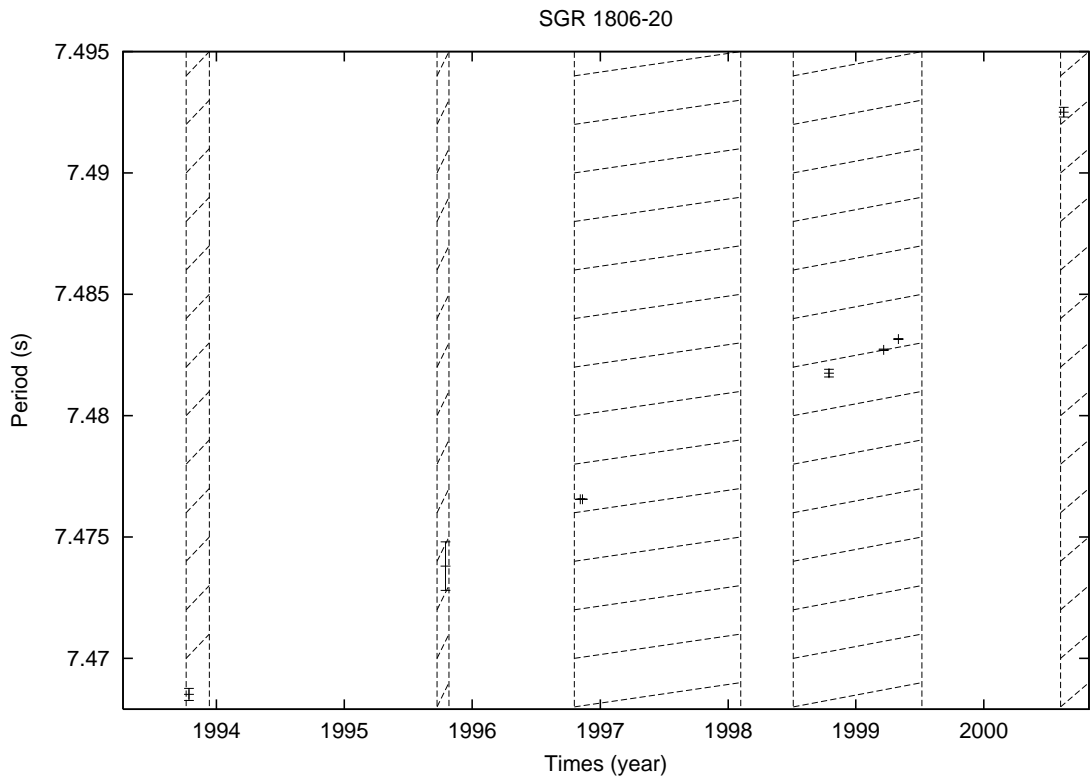,width=15cm,angle=-90}}
\caption{Period evolution of SGR 1806-20. Two consequitive dashed-lines
represent the bursting intervals detected by BATSE}
\end{figure*}

\begin{figure*}
\centerline{\psfig{file=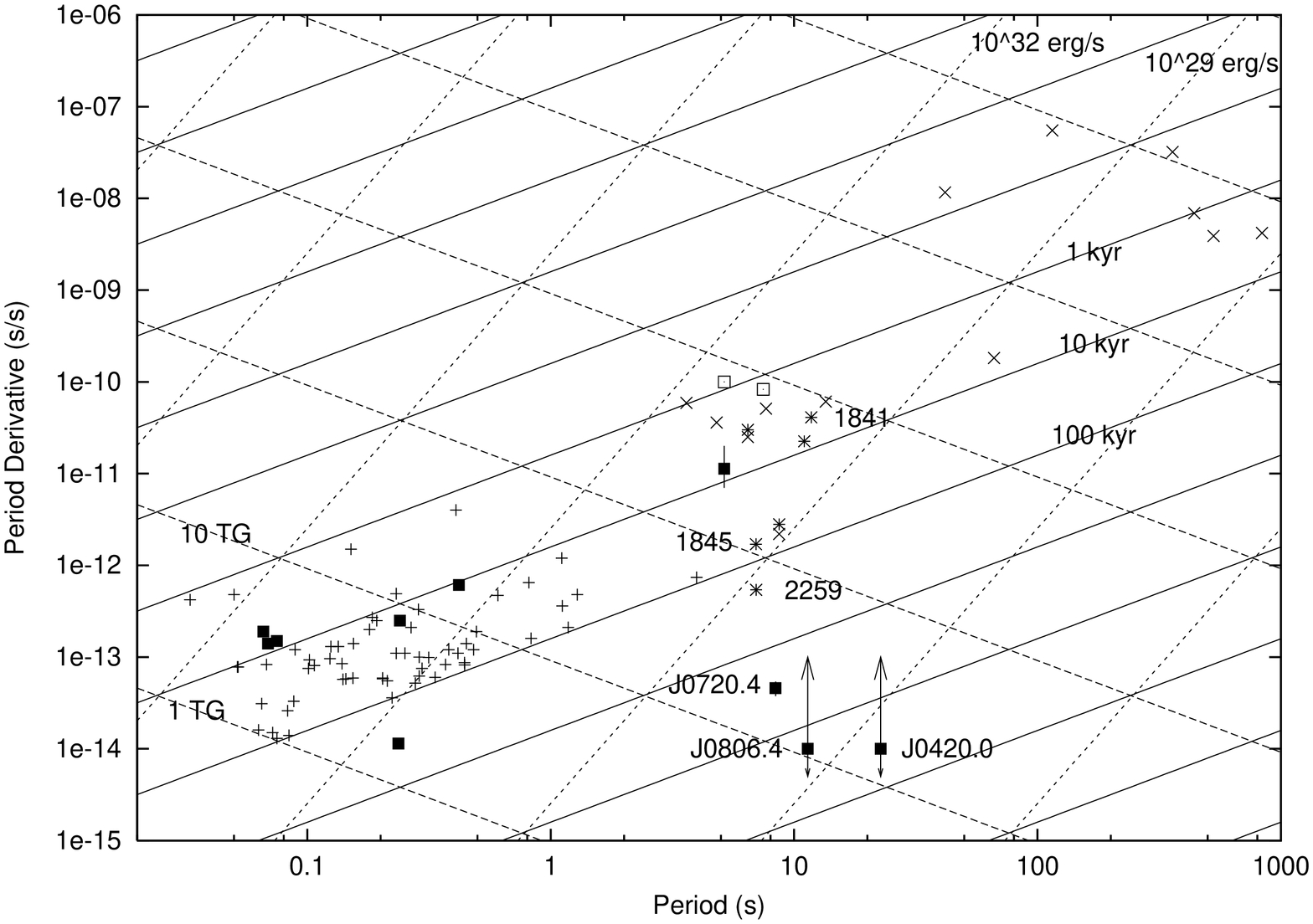,width=16cm,angle=-90}}
{Figure 8. P-\.{P} Diagram for PSRs, AXPs, SGRs, XBs and DRQNSs. +
symbol
represents PSR, * represents AXP, boxes represent SGRs, filled boxes 
DRQNSs and x represents XB. 
For DRQNS RX J0002+6246  and
1E1207.4-5209
period derivative is calculated using its period value, and
age of the associated SNR}
\end{figure*}  

\begin{figure}
\centerline{\psfig{file=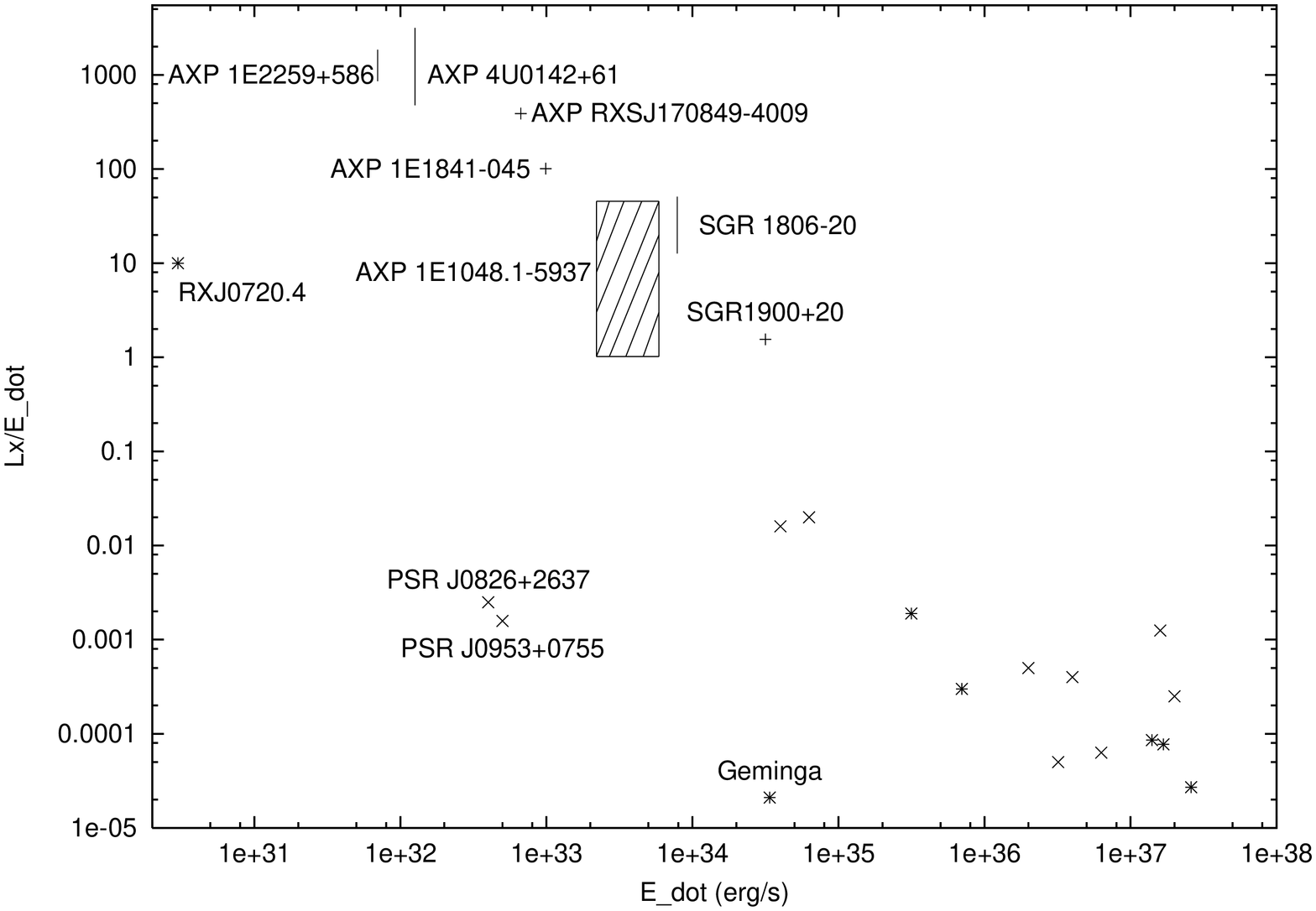,width=16cm,angle=-90}}
{Figure 9. Ratio of total X-ray luminosity L$_x$ to rotational energy 
loss rate \.{E} versus \.{E} for PSRs, DRQNSs, AXPs and SGRs. PSRs are 
represented by 
*'s, DRQNSs are represented by x's and AXPs/SGRs are represented by +'s.}
\end{figure}

\end{document}